\documentclass[a4paper,fleqn,usenatbib]{mnras}

\usepackage{mathptmx}
\usepackage[T1]{fontenc}
\usepackage{ae,aecompl}
\usepackage[normalem]{ulem}

\usepackage{graphicx}	% Including figure files
\usepackage{amsmath}	% Advanced maths commands
\usepackage{amssymb}
\usepackage{subfigure}% Extra maths symbols

\def\kms{km\,s$^{-1}$}

\title[The OH Megamaser galaxy IRAS03056]{Star formation and gas inflows in the OH Megamaser galaxy IRAS03056+2034  }

\author[C. Hekatelyne et al.]{C. Hekatelyne,$^{1}$\thanks{E-mail: hekatelyne.carpes@gmail.com}
Rogemar A. Riffel,$^{1}$ Dinalva Sales,$^2$  Andrew Robinson,$^3$  
\newauthor Thaisa Storchi-Bergmann,$^4$ Preeti Kharb,$^5$ Jack Gallimore,$^6$ Stefi Baum,$^{7,8}$
\newauthor   Christopher  O'Dea$^{7,9}$ \\
$^{1}$ Departamento de F\'isica, CCNE, Universidade Federal de Santa Maria, 97105-900, Santa Maria, RS, Brazil\\
$^{2}$ Instituto de Matem\'atica, Estat\'istica e F\'isica, Universidade Federal do Rio Grande, Rio Grande 96203-900, Brazil \\
$^{3}$ School of Physics and Astronomy, Rochester Institute of Technology, 84 Lomb Memorial Drive, Rochester, NY 14623, USA \\
$^4$ Departamento de Astronomia, Universidade Federal do Rio Grande do Sul. 9500 Bento Gon\c{c}alves, Porto Alegre, 91501-970, Brazil \\
$^5$ National Centre for Radio Astrophysics, Tata Institute of Fundamental Research, S. P. Pune University Campus, Post Bag 3,\\ Ganeshkhind, Pune 411 007, India \\
$^6$ Department of Physics, Bucknell University, Lewisburg, PA 17837, USA \\
$^7$ Department of Physics and Astronomy, University of Manitoba, Winnipeg, MB, R3T 2N2, Canada\\
$^8$ Center for Imaging Science, Rochester Institute of Technology, 84 Lomb Memorial Dr., Rochester, NY 14623, USA.\\
$^9$ School of Physics \& Astronomy, Rochester Institute of Technology, 84 Lomb Memorial Dr., Rochester, NY 14623, USA. }

\date{Accepted XXX. Received YYY; in original form ZZZ}

\pubyear{2017}

\begin{document}
\label{firstpage}
\pagerange{\pageref{firstpage}--\pageref{lastpage}}
\maketitle

\begin{abstract}

We have obtained observations of the OH Megamaser galaxy IRAS03056+0234 using Gemini Multi-Object Spectrograph (GMOS) Integral Field Unit (IFU), Very Large Array (VLA) and Hubble Space Telescope (HST). The HST data reveals spiral arms containing knots of emission associated to star forming regions. The GMOS-IFU data cover the spectral range of 4500 to 7500 ~\AA ~at a velocity resolution of 90 km s$^{-1}$ and spatial resolution of 506 pc. The emission-line flux distributions reveal a ring of star forming regions with radius of 786 pc centred at the nucleus of the galaxy, with an ionized gas mass of 1.2$\times$ 10$^{8}$M$_{\odot}$, an ionizing photon luminosity of log Q[H$^{+}$]=53.8 and a star formation rate of 4.9 M$_{\odot}$ yr$^{-1}$. The emission-line ratios and radio emission suggest that the gas at the nuclear region is excited by both starburst activity and an active galactic nucleus. The gas velocity fields are partially reproduced by rotation in the galactic plane, but show, in addition, excess redshifts to the east of the nucleus, consistent with gas inflows towards the nucleus, with  velocity of $\sim$45\,km s$^{-1}$ and a mass inflow rate of $\sim$7.7$\times$10$^{-3}$ M$_{\odot}$ yr$^{-1}$. 

\end{abstract}


%%%%%%%%%%%%%%%%%%%%%%%%%%%%%%%%%%%%%%%%%%%%%%%%%%

%%%%%%%%%%%%%%%%% BODY OF PAPER %%%%%%%%%%%%%%%%%%

\section{Introduction}

OH Megamasers (hereafter OHMs) are powerful extragalactic masers in which the emission occurs predominantly in the 1667/1665 MHz lines
with typical luminosities of about 10$^3$\,L$_{\odot}$. The isotropic luminosity of OHM's is ~10$^{8}$ times greater 
than the luminosity of the OH masers in the Milk Way. The megamasers can be explained on the basis of amplification of the nuclear radio continuum by foreground molecular material with inverted level populations arising from some pumping source \citep[e.g.][]{baan98,lo2005,chen2007}.

In general, OHMs have been observed in (Ultra) Luminous Infra-Red Galaxies [(U)LIRGs] with infra-red luminosities of L$_{\rm IR}$ >= 10$^{11}$ L$_{\odot}$. These merging systems fulfill all the requirements for producing OHM emission. The merger interaction concentrates molecular gas in the galaxy nuclei, creates strong dust emission from the
starburst (SB) and Active Galactic Nuclei (AGN) activity, and produces radio continuum emission from the AGN or SB \citep[e.g.][]{henkel87,Darling2000,chen2007}.

It is not well understood if the mechanism of ionization in the systems that host OHM is dominated by star formation or  AGN activity, in the sense that the hosts of OHM emission usually present features of both phenomena in their spectra. A possible explanation for these features is that they originate in a central AGN, contaminated by emission of circumnuclear star-forming regions, as the angular resolution of the observations usually corresponds to a few kiloparsecs at the galaxies. Alternatively, the OHM galaxies could represent a transition stage between a starburst and the emergence of an AGN, as suggested by \citet{Darling2006}.

In an effort to investigate the nature of the ionization in OHM galaxies we have been performing a multi-wavelength study of a sample of these galaxies \citep{Sales2015,heka18}. In this paper we present Gemini Multi-Object Spectrograph (GMOS) Integral Field Unit (IFU) observations, Very Large Array (VLA) continuum data and Hubble Space Telescope (HST) narrow and broad band images of the galaxy IRAS03056+2034, which is a LIRG that hosts OHM emission. This target is part of a sample including another 14 OHM galaxies, for which we have the same combination of HST and VLA data. The targets selected for IFU observations were chosen based on the morphology revealed by the HST images. This paper is part of a series based on multi-wavelength observations with the aim of studying the gas kinematics and excitation of OHM galaxies.

In a previous paper \citep{heka18} we mapped the eastern nucleus of the OHM galaxy  IRASF23199+0123 using HST, VLA and GMOS data. We were able to conclude that the object is an interacting pair with a tail connecting the two galaxies and detected two OH maser sources associated to the eastern member. Moreover, we discovered a Seyfert 1 nucleus in the eastern member of the pair, via detection of an unresolved broad double peaked component in the H$_\alpha$ emission-line. In addition, the masing sources were observed in the vicinity of a region of enhanced velocity dispersion and higher line ratios, suggesting that they are associated with shocks driven by AGN outflows. These results  suggest that the OH megamaser emission in IRASF23199+0123 is associated to AGN activity. \citet{Sales2015} presented a multi-wavelength study of the OH megamaser galaxy IRAS16399--0937 using  HST, VLA, 2MASS, Herschel and Spitzer data. This galaxy has two nucleus separated by 3.4~kpc and its infrared spectrum is dominated by strong polycyclic aromatic hydrocarbon, but the northern nucleus shows in addition deep silicate and molecular absorption features. The analysis of the spectral energy distribution reveals that the northern nucleus contains an AGN with bolometric luminosity of $10^{44}$\,erg\,s$^{-1}$. 

%The evidence of this multiwavelength study allowed us to conclude that the OH megamaser sources are associated with the AGN rather than star formation.

IRAS03056+2034 (hereafter IRAS03056) is a spiral galaxy (SB(rs)B \citep{Vaucouleurs91}) that presents strong OHM emission. This detection was obtained with the Nancay radio telescope in 1990, indicating F$_{IR}$ luminosity of 15$\times$10$^{10}$ L$_{\odot}$ \citep{Bottinelli90}. \citet{baan98} used spectroscopic data obtained with the 200 inch Hale telescope at the Palomar Observatory in order to determine the optical classifications of 42 OH Megamaser galaxies, based on line ratios, and classified IRAS03056 as a starburst galaxy. We adopt the distance of 116 Mpc as derived by \citet{Theureau07} from the Tully-Fisher relation. %1$^{\prime\prime}$ corresponds to $\sim$560\,pc at the galaxy.

We have obtained GMOS-IFU data covering the central region of IRAS03056 in order to map the distribution and kinematics of the emitting gas and investigate the origin of the line emission in the central region of this object.
%Hekatelyne et al, 2017 performed a similar investigation for another OH Megamaser galaxy, revealing a Seyfert 2 nucleus and an outflow related to it. This paper is organized as follows.
This paper is organized as follows. The observations and the data reduction procedure are described in Sec.~2, the emission-line flux distributions, emission-line ratio and kinematics maps obtained from GMOS data, as well as the HST and VLA images are presented in Sec.~3. These results are discussed in Sec.~4 and then in Sec.~5 we present the final remarks. 

\section{Observations and data reduction}

\subsection{HST images}

The HST images of IRAS03056 were obtained with the Advanced Camera for Surveys (ACS). The acquired images consist of continuum and emission line imaging of a sample of 15 OHM galaxies (Program id 11604; PI: D.J. Axon). The observations were done with the wide-field channel (WFC) using broad (F814W), narrow (FR656N) and medium-band (FR914M) filters. The total integration times were 600\,sec for the broad band filter, 200\,sec for the medium-band and 600\,sec in the narrow-band filter which contains the H$\alpha$ and [N\,{\sc ii}] lines. 

The images were processed using IRAF packages \citep{tody86,tody93}. First of all, the cosmic rays were removed from the images using the {\it lacos im} task \citep{Vandokkum2001}. In order to build a continuum-free H$\alpha+$[N{\sc ii}] image of IRAS03056 we estimate the count rate for foreground stars considering the medium and narrow band images. This procedure allowed us to define a mean scaling factor that was applied to the medium-band image. It provided us a scaled image that was subtracted from the narrow band image (Sales et al., in preparation). 

Finally, the continuum-free image was inspected to certify that the residuals at the positions of the foreground stars are negligible. This procedure results in typical uncertainties of 5-10 \% in flux  \citep[see][]{Hoopes1999,Rossa2000,Rossa2003}.

\subsection{VLA Radio Continuum data}

We reduced archival VLA A-array data at 1.425 GHz from the project AB660. These data were acquired on 14 December 1992. The data were reduced following standard procedures in AIPS. The final image of IRAS03056 was created after a couple of phase-only and phase+amplitude self-calibrations, using the AIPS tasks CALIB and IMAGR iteratively. The r.m.s. noise in the image is  $\sim7\times10^{-5}$~Jy~beam$^{-1}$. The restoring beam is $1.49\times1.38$~arcsec at a PA = $-35\degr$. The peak intensity of the compact radio core seen in the image is $\sim11$~mJy~beam$^{-1}$. An image at 4.86 GHz from the project AB660 at a resolution of $0.38\times0.34$~arcsec (beam PA=$-34.5\degr$) was also available in the NRAO image archive. A point source of peak intensity $\sim1.7$~mJy~beam$^{-1}$ is visible in this image; the r.m.s. noise is $\sim8.7\times10^{-5}$~Jy~beam$^{-1}$. 

%We created a (1.425 - 4.86) GHz spectral index image of IRAS03056 and found that the core has a steep spectral index of $-1.02\pm0.09$; this is consistent with optically thin synchrotron emission that could arise in an AGN jet.

%The Very Large Array (VLA) data for IRAS03056 were obtained on \textcolor{red}{ XXXX. The observations comprises the ?-band.}

%The reduction process was performed using the VLA pipeline in CASA \citep{McMullin2007}. The main procedures include the initial data flagging and phase, flux and bandpass calibrations. The continuum image was generated using multi-frequency synthesis \citep[e.g.][]{Conway90,Rau2011} with natural weighting and deconvolved using the Cotton-Schwab variant of the CLEAN algorithm \citep{Schwab84}.

%The imaging included simultaneous deconvolution of neighboring radio sources within the primary beam. Also, three rounds of phase-only self-calibration based on CLEAN models for the radio continuum (self-calibration is reviewed by \citet{Pearson84}).

%\textcolor{red}{For the L-band continuum image, the restoring beam is XXX, PA=XXX, and the background rms is XXX~mJy~beam$^{-1}$.}

\subsection{GMOS-IFU data}

Optical spectroscopic data for IRAS03056 was obtained at the Gemini North telescope, using the Gemini Multi-Object Spectrograph Integral Field Unit \citep[GMOS-IFU,][]{allington-smith02,hook04}. The observations were performed in 2013, October, November and December (Gemini project GN-2013B-Q-97 -- PI: D. Sales).
The observations were carried out using the B600 grating in combination with the G5307 filter, with the major axis of the IFU oriented along position angle $PA=120^\circ$, approximately along the major axis of the galaxy.

The total integration time was 12\,000\,sec divided into 10 individual exposures of 1\,200\,sec each. The one slit mode  of GMOS IFU was used, resulting in a total angular coverage of 5\farcs0$\times$3\farcs5, and a spectral range that includes the H$_{\alpha}$, [N\,{\sc ii}]$\lambda$6583, [S\,{\sc ii}]$\lambda$6717, H$\beta$, [O\,{\sc iii}]$\lambda$5007 and [O\,{\sc i}]$\lambda$6300 emission-lines.

In order to process the data we followed the standard steps for spectroscopic data reduction using GEMINI package routines of IRAF \citep{lena2014}. The basic steps of data reduction comprise bias level subtraction, flatfielding, trimming and wavelength calibration. We used the CuAr arc lamps as reference in order to apply the wavelength calibration to the data and subtracted the sky emission contamination. In order to apply the flux calibration we used a sensitivity function that was generated from a spectrum of the BD+28 4211 photometric standard star, observed in the same night of the galaxy exposures.

After flux calibration we created datacubes for each exposure at a sampling of 0\farcs1$\times$0\farcs1. These datacubes were median combined using IRAF {\it gemcombine} task resulting in the final data cube for the object. In the mosaicking process, we used as a reference the peak of the continuum emission and used the  {\it sigclip} algorithm to remove bad pixels. 

 The adopted GMOS configuration resulted in a spectral resolution of 1.8 \AA, as obtained from the Full-width at half maximum (FWHM) of CuAr arc lamp lines used to perform the wavelength calibration, corresponding to 90 km s$^{-1}$. The angular resolution is 0.9$^{\prime\prime}$, as measured from the FWHM of field stars in the acquisition image. This corresponds to 506 pc at the galaxy.

As the final cube presented unwanted noise, we performed a spatial filtering using the IDL routine $bandpass_-filter.pro$\footnote{The routine is available at $https://www.harrisgeospatial.com/docs/bandpass_-filter.html$}, which allows the choice of the cut-off frequency ($\nu$) and the order of the filter $n$. The filtering process does not change the angular resolution of the data and all measurements presented in the forthcoming sections were done using the filtered cube.

\section{Results}

\begin{figure*}
	\includegraphics[width=0.95\textwidth]{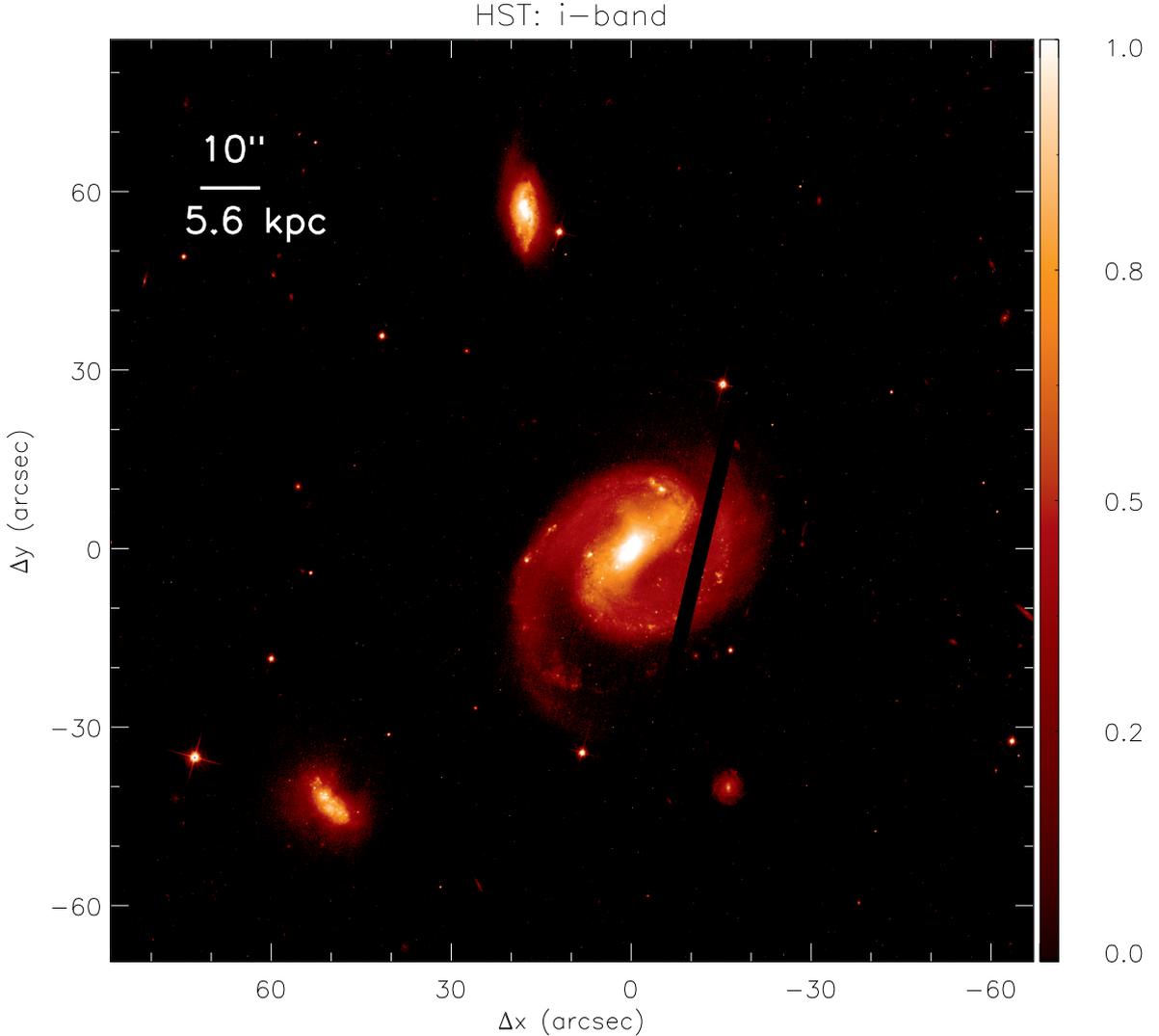}
    \caption{Large scale i-band image of IRAS03056 obtained with HST.}
    \label{fig:merger}
\end{figure*}

\subsection{Large-scale structure and merger stage}

IRAS03056+2034 has been spectroscopically classified as a starburst \citep{baan98} with infrared and OH maser luminosities of $L_{\rm IR}=1.6\times10^{11}$\,L$_{\odot}$ and log\,L$_{OH}=1.3$\,L$_{\odot}$ \citep{baan08,Kandalian1996}.

Figure~\ref{fig:merger} shows the large scale i-band image of IRAS03056 obtained with HST. This image shows a  barred spiral galaxy  with two  symmetric arms.  Two smaller companion galaxies at projected distances of roughly 31\,kpc and 34.1\,kpc from the nucleus of IRAS03056 are seen to the north and to south-east, respectively.

\citet{Haan2011} studied a sample of 73 nearby ($0.01 < z < 0.05$ ) LIRGs and classified the objects into six different merger stages, based on H-band HST images.
% (1) separated galaxies, with symmetric disks and no tidal tails, (2) progenitor galaxies distinguishable with disks asymmetric or amorphous and/or tidal tails, (3) two nuclei in common envelope,  (4) double nuclei plus tidal tail, (5) single or obscured nucleus with disturbed central  morphology and short faint tails. 
Using theirs classification scheme,  IRAS03056 is classified as type 1 (separated galaxies)  and this group constitutes only 8.3\% of their sample. 
%These studies lead us conclude that IRAS03056 is an unusual in terms of its merger stage among (U)LIRGs. 

Our large scale images exhibit a scenario that, in general, IRAS03056 is a fairly typical example of a LIRG that has low infrared luminosity (L$_{IR}=1.6\times10^{11}$L$_{\odot}$) 
with a wide nuclear projected separation between the nuclear components, however, presenting composite (starburst+AGN) spectra \citep[see Fig. 12 and Tab. 6 of][]{Yuan2010}. In terms of star-formation and molecular gas content, it is also important to note that IRAS03056 is widely similar 
to (U)LIRGs lying at the high end of the linear relation between the surface density of dense molecular gas (as traced by HCN) and the  surface density of star-formation rate (SFR) with a positive correlation coefficient  \citep[see Fig. 5 and Tab. 2 of][]{Liu2015}.

%In previous studies we used multi-wavelength observations of other two OHM galaxies, IRAS16399-0937 \citep{Sales2015} and IRASF23199+0123 \citep{heka18}, 
The two OHM galaxies previously studied by our group \citep{Sales2015,heka18} show a more advanced merger stage than IRAS03056, presenting  ``close binary"  nuclei. In addition, the previously studied galaxies are slightly more luminous (log L$_{\rm IR} \approx$ 11.5 -- 11.6 L$_{\odot}$) than IRAS03056.

%(i.e., early stage merger)  unusual in terms of its merger stage among (U)LIRGs and slightly brighter (L$_{IR} \approx$ 11.5 -- 11.6 L$_{\odot}$) than IRAS03056 \citep{Sales2015,heka18}.

%\subsection{Emission-line profile fitting}

\begin{figure}
	\includegraphics[width=0.45\textwidth]{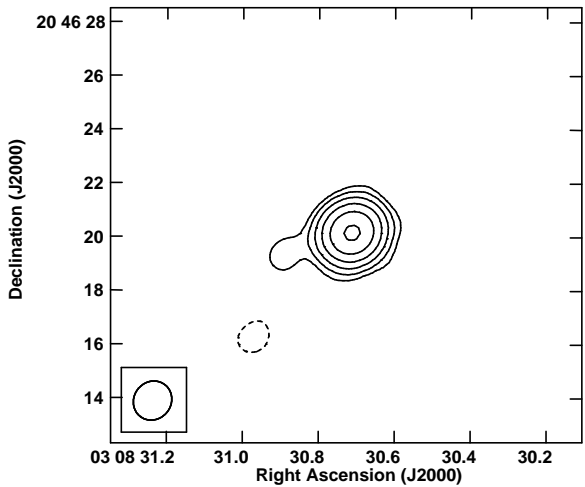}
    \caption{VLA radio continuum image of IRAS03056 at 20 cm. The contour levels are in percentage of the the peak intensity (=10.9 mJy~beam$^{-1}$) and increase in steps of two, with the lowest contour levels being $\pm2.8$\%.}
    \label{fig:radio}
\end{figure}

\subsection{VLA and HST images}

Figure~\ref{fig:radio} shows the 20\,cm continuum image of IRAS03056 obtained with VLA. A compact radio source is detected at the nucleus of the galaxy and no extended emission is seen. We created a (1.425 - 4.86) GHz spectral index image of IRAS03056 and found that the core has a steep spectral index of $-1.02\pm0.09$; this is consistent with optically thin synchrotron emission that could arise in an AGN jet  \citep{pacholczyk70}.
Moreover, we estimated the brightness temperature ($T_B$) of the 1.4 GHz radio core by using the total flux density (=11.7 mJy) and the beam-deconvolved size of the core (=0.51" x 0.24"), as derived from the AIPS Gaussian-fitting task JMFIT, and the $T_B$ relation for an unresolved component from \citet{Ulvestad2005}. This turned out to be = $1.8\times10^5$~K, supporting an AGN-related origin \citep[e.g.][]{berton2018}.

\begin{figure*}
	\includegraphics[width=1.9\columnwidth]{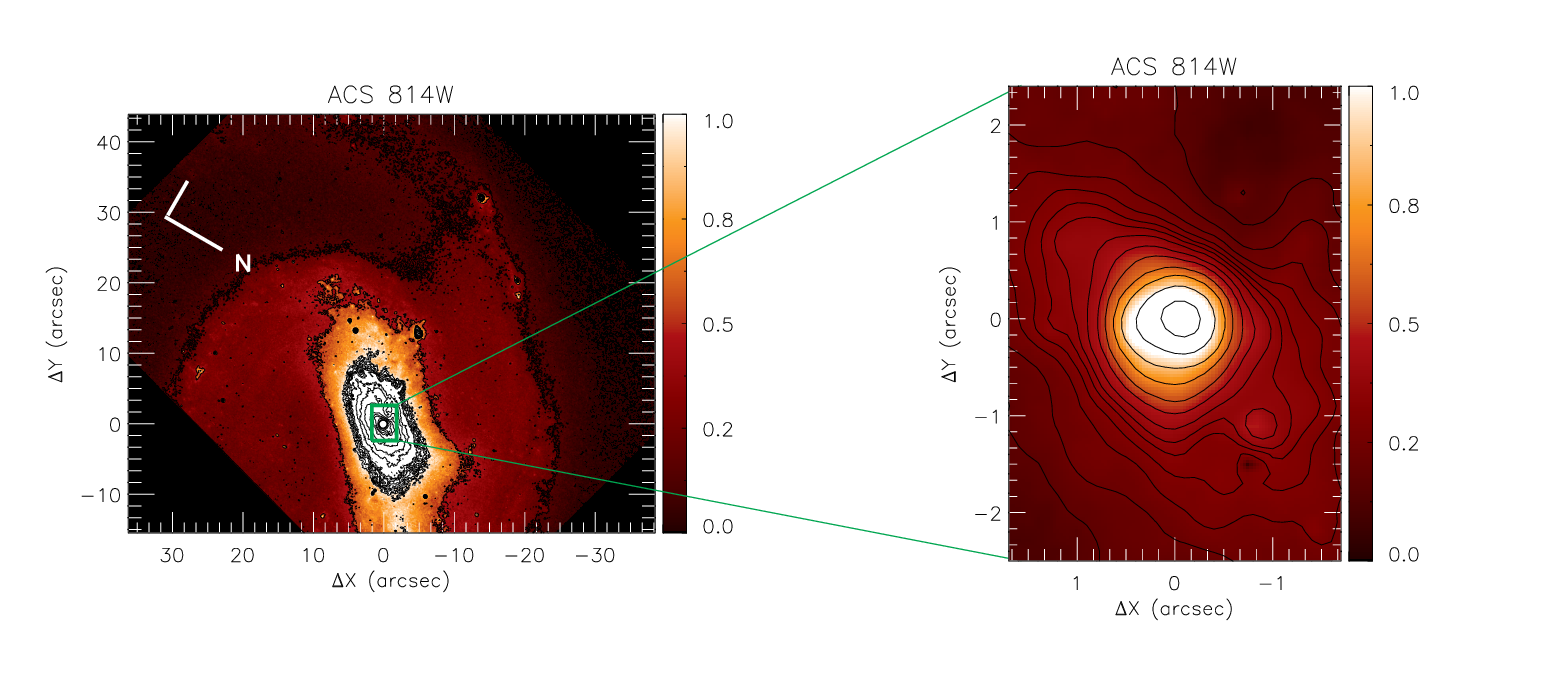}
	\includegraphics[width=1.9\columnwidth]{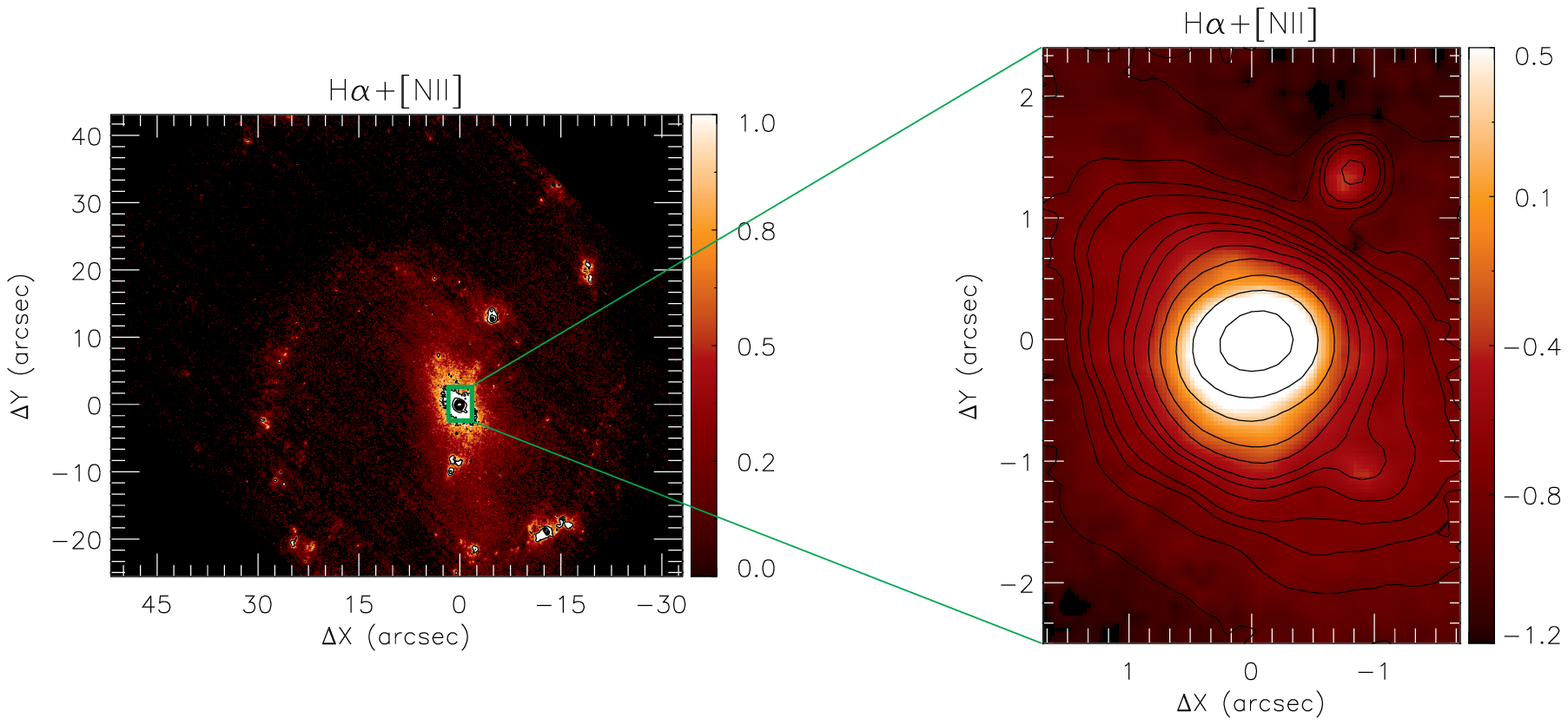}
    \caption{HST images of IRAS03056. Top panels: Left -- Large-scale image (ACS/HST F814W - i band).  Right -- Zoom of i band image for the  region observed with GMOS-IFU. Bottom panels: Left -- Large-scale continuum free H$\alpha$+[N\,{\sc ii}] image. Right -- Zoom at the region observed with GMOS IFU. The green boxes show the GMOS IFU FoV (3\farcs5$\times$5\farcs5) and the color bars show the fluxes in arbitrary units.}
%{\bf Bottom panels: Left -- Structure map. The green box represents the IFU field. Right -- Expanded image of the IFU field.}}
    \label{fig:hst-images}
\end{figure*}

The HST images of IRAS03056 are presented in Figure \ref{fig:hst-images}. The left panels present the broad-band continuum (F814W) image (top panel) and the narrow-band [N\,{\sc ii}]+H$\alpha$ image (bottom panel) of the inner 20$\times$20 arcsec$^{2}$ of IRAS03056. The green boxes  represent the field-of-view (FoV) of the GMOS-IFU data. The right panels shows a zoom of the central region of the HST images, corresponding to the GMOS FoV. The HST images were rotated to the same orientation of the GMOS-IFU data.

The HST continuum image  shows the most elongated emission along PA$\sim$100/280$^{\circ}$ and  presents a structure that seems to be associated with a spiral arm  seen to the northeast of the nucleus. Moreover, the zoomed image (top-right panel) shows a strip of emission that extends from 1$^{\prime\prime}$ north to 1$^{\prime\prime}$ south of the nucleus.

%The bottom left panel of Fig.~\ref{fig:hst-images} shows the continuum-free H$\alpha$+[N\,{\sc ii}] narrow-band HST image of the galaxy
 The H$\alpha$+[N\,{\sc ii}] flux distribution is similar to that in the continuum but shows more clearly the presence of spiral arms, one to the west and another to the southeast of the nucleus. At the central region (bottom-right panel) unresolved knots of emission are seen in both H$\alpha$+[N\,{\sc ii}] and i band images, one at 1\farcs5 north and another at 1\farcs5  east of the nucleus. %This knots of emission seems to be associated the enhanced continuum emission seen at the  F814W image (top-right panel). 
 
 %The HST images are similar at the highest flux intensities, presenting elongated structure along PA 100/280$^{\circ}$. In addiction, the H$\alpha$+[N\,{\sc ii}] image clearly evince several knots of star forming regions located at the spiral arms.

%Figure \ref{fig:radio} presents the radio continuum images at 20 cm. The radio image shows a compact structure, elongated in the North-South direction. (INFLOW?)
%\textcolor{red}{Descrever melhor. As imagens são parecidas nos níveis mais intensos de fluxos, com uma estrutura mais estendida alongo do PA=100/280. In addition, a imagem do NII+HA mostra claramente os braços espirais com several knots of star forming regions...}

%\textcolor{red}{Nada foi falado sobre a imagem do VLA, como sugere o título da seção. Descrevê-la como compacta, ligeiramente mais alongada na direçao N-S...seria coincidencia que o inflow é visto para N}

\subsection{Emission-line flux distributions}\label{flux-res}

\begin{figure*}
    \centering
    \includegraphics[width=0.95\textwidth]{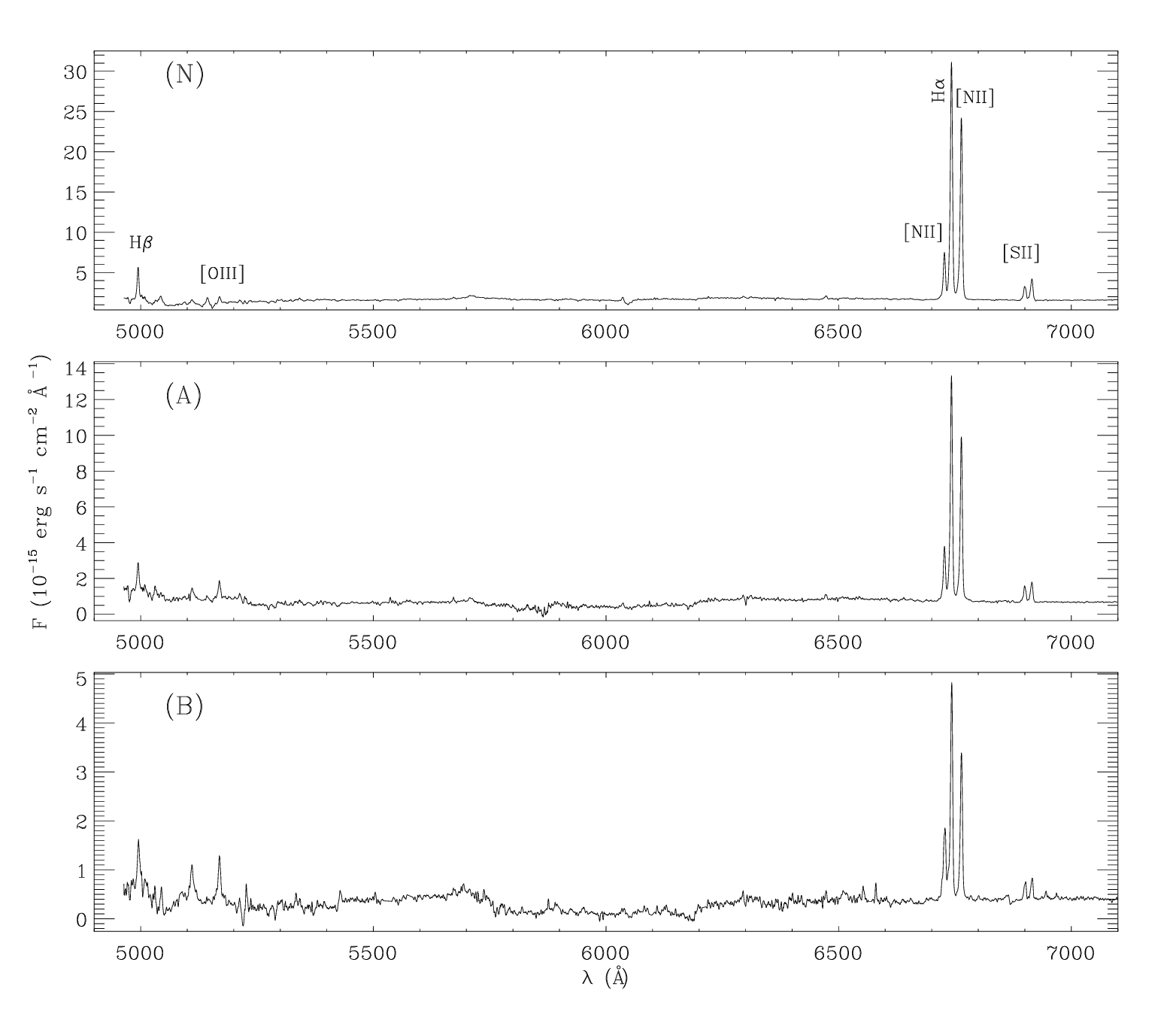}
    \caption{Examples of spectra for IRAS03056 obtained within a circular aperture of 0\farcs45 for the nucleus (top), region A (middle) and B (bottom) labeled in Fig.~\ref{fig:fluxo}. The strongest emission lines are identified in the nuclear spectrum.}
    \label{fig:espectros}
\end{figure*}

\begin{figure*}
	\includegraphics[width=2.2\columnwidth]{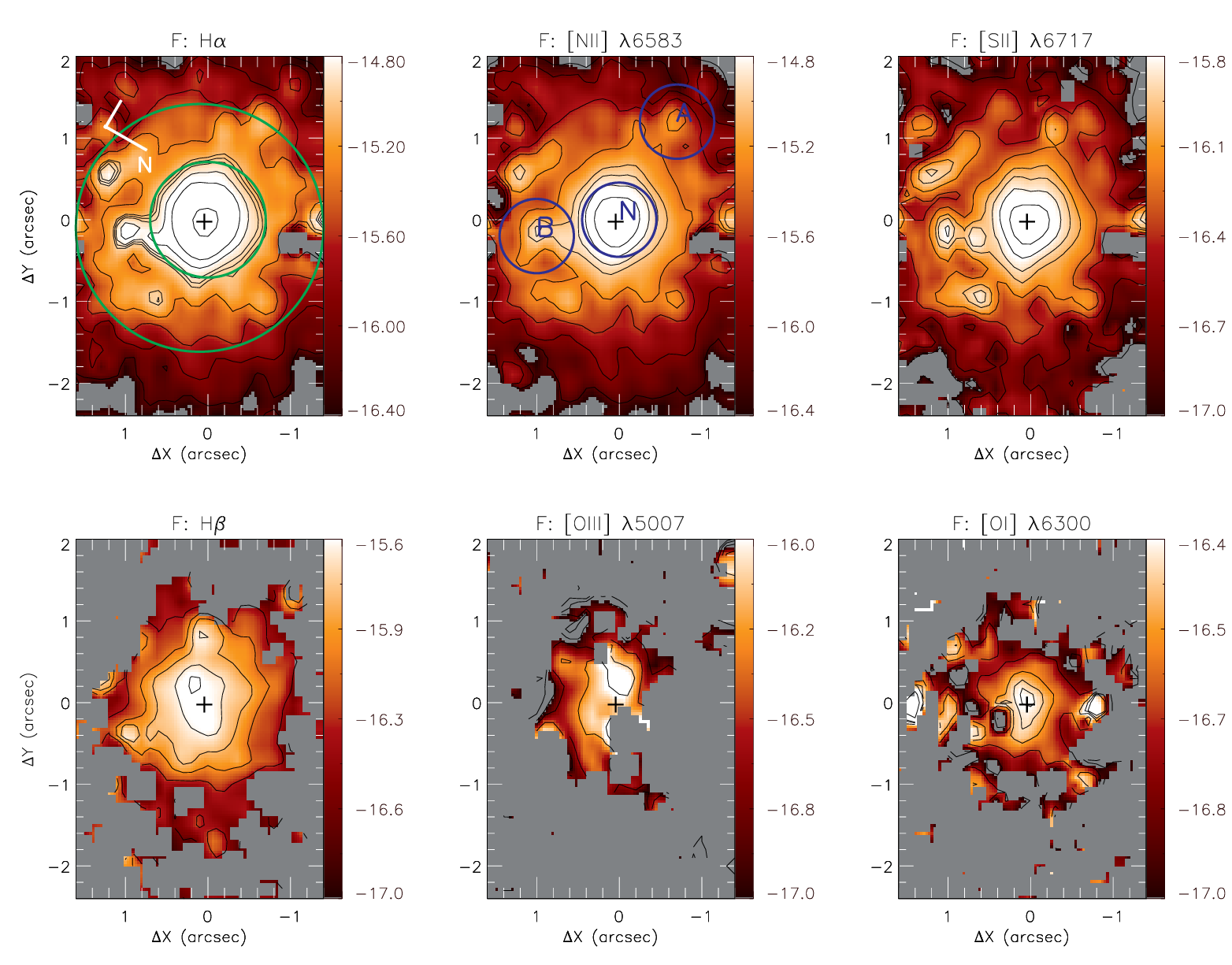}
\caption{Top panels: flux maps for H$\alpha$ (left), [N\,{\sc ii}]$\lambda$6583 (centre) and [S\,{\sc ii}]$\lambda$6717 (left) emission-lines of IRAS03056. The green circle delimits a ring where we have extracted a spectra in order to characterize the star formation. The blue circles labeled as N, A and B represent the circular regions where we extracted the spectra shown in Fig.~\ref{fig:espectros}. Bottom panels: flux maps for H$\beta$ (left), [O\,{\sc iii}]$\lambda$5007 (centre) and [O\,{\sc i}]$\lambda$ 6300 (right). The central crosses in all maps mark the position of the nucleus and grey regions represent masked locations, where the signal-to-noise was not high enough to obtain reliable fits of the emission-line profiles or locations with no line detection. The color bars show the fluxes in logarithmic units of erg\,s$^{-1}$cm$^{-s}$ and the grey regions represent masked locations where the signal-to-noise ratio was not high enough to obtain good fits of the emission-line profiles. These locations present flux uncertainty larger than 30\%.}
\label{fig:fluxo}
\end{figure*}

%\textcolor{red}{Essa seção está deslocada. Acho melhor colocar o seu conteúdo no começo da seção Emission-line flux distributions}
Figure \ref{fig:espectros} shows examples of IRAS03056 spectra obtained from the GMOS-IFU datacube for the three locations indicated as blue circles in the top-middle panel of Fig.~\ref{fig:fluxo} and labeled as N (nucleus), A (1\farcs5 east) and B (1\farcs0 southwest). These  spectra were obtained by  integrating the fluxes within circular apertures of 0\farcs45 radius. The strongest emission lines are identified in the nuclear spectrum.

Aiming to map emission-line flux distributions, line-of-sight velocity ($V_{LOS}$) and velocity dispersion ($\sigma$) of the emitting gas, we have fitted the emission-line profiles of H$\alpha$, [N\,{\sc ii}]$\lambda$6548,6583, H$\beta$, [S\,{\sc ii}]$\lambda$6717, [O\,{\sc i}]$\lambda$6300 and [O\,{\sc iii}]$\lambda$5007 by Gaussian curves. The fitting procedure was performed using modified versions of the line-PROfile FITting ({\sc profit}) routine \citep{profit}. This routine performs the modeling of the observed emission-line profile using the MPFITFUN routine \citep{mark09}, via a non-linear least-squares fit. The outputs of the routine are the emission-line flux, the centroid velocity, the velocity dispersion and their corresponding uncertainties for each emission line. 

The fitting process for the [N\,{\sc ii}]$\lambda$6548,6583+H$\alpha$ complex was performed simultaneously, considering one Gaussian per line.
During the fit, we kept tied the kinematics ($V_{LOS}$ and $\sigma$) of the [N\,{\sc ii}] and fixed the [N\,{\sc ii}]$\lambda$6583/[N\,{\sc ii}]$\lambda$6548 intensity ratio to its theoretical value \citep[3,][]{osterbrock}. The underlying continuum was fitted by a linear equation, constrained by the adjacent continuum regions.

Figure \ref{fig:fluxo} presents the flux distributions for H$\alpha$, [N\,{\sc ii}]$\lambda$6583, [S\,{\sc ii}]$\lambda$6717, H$\beta$, [O\,{\sc iii}]$\lambda$5007 and [O\,{\sc i}]$\lambda$6300 emission-line. The color bars show the flux in logarithmic units of erg s$^{-1}$cm$^2$ and the grey regions represent masked locations where the signal-to-noise ratio was not high enough to obtain good fits of the emission-line profiles. These locations present flux uncertainty larger than 30\%. The central crosses mark the location of the nucleus, defined as the position of the peak of continuum emission. H$\alpha$, [N\,{\sc ii}]$\lambda$6583 and [S\,{\sc ii}]$\lambda$6717 emission-lines flux distributions are similar presenting extended emission over the whole GMOS FoV. One can notice the presence of unresolved knots of emission surrounding the nucleus at $\sim$1$^{\prime\prime}$ from it. 

%The region containing these knots of emission is delimited by the green circles shown in the H$_{\alpha}$ flux map.

%\textcolor{red}{Fico na dúvida sobre pq estes knots não aparecem tão claramente nas imagens do HST. Deveriam aparecer mais claramente por causa da melhor resolução === not resolved => unresolved}

The bottom panels of Fig.~\ref{fig:fluxo} show the flux distribution maps for H$\beta$, [O\,{\sc iii}]$\lambda$5007 and [O\,{\sc i}]$\lambda$6300, which are generally detected only within a smaller region around the nucleus. As for H$\alpha$, the H$\beta$ flux map shows some unresolved knots of emission at $\sim$1$^{\prime\prime}$ away from the nucleus and faint extended emission is seen to up to 2$^{\prime\prime}$ from it. The  [O\,{\sc iii}]$\lambda$5007 emission is detected only very close to the nucleus, at distances smaller than 1$^{\prime\prime}$. The [O\,{\sc i}]$\lambda$6300 emission is also mostly concentrated within $\sim$1$^{\prime\prime}$ of the nucleus.

%Most of the  [O\,{\sc i}]$\lambda$6300 emission is seen at similar distances to those of the [O\,{\sc iii}]$\lambda$5007 flux map, although some emission is seen at distances of up to  2$^{\prime\prime}$ from the nucleus.

%\textcolor{red}{dá para descrever mais os mapas...até onde se estendem? Algum problema nos espectros na região do OIII próximo do núcleo a W, que faz com que não tenha medida la? Se sim, dá para esplicar...}

\subsection{Line-Ratio maps}

%\textcolor{red}{esta seçao deve vir antes da cinemática...começar a seção dizendo pra que servem estas razões, sem especificar o que cada valor representa=> isso fica para as discussões}

\begin{figure*}
	\includegraphics[width=\textwidth]{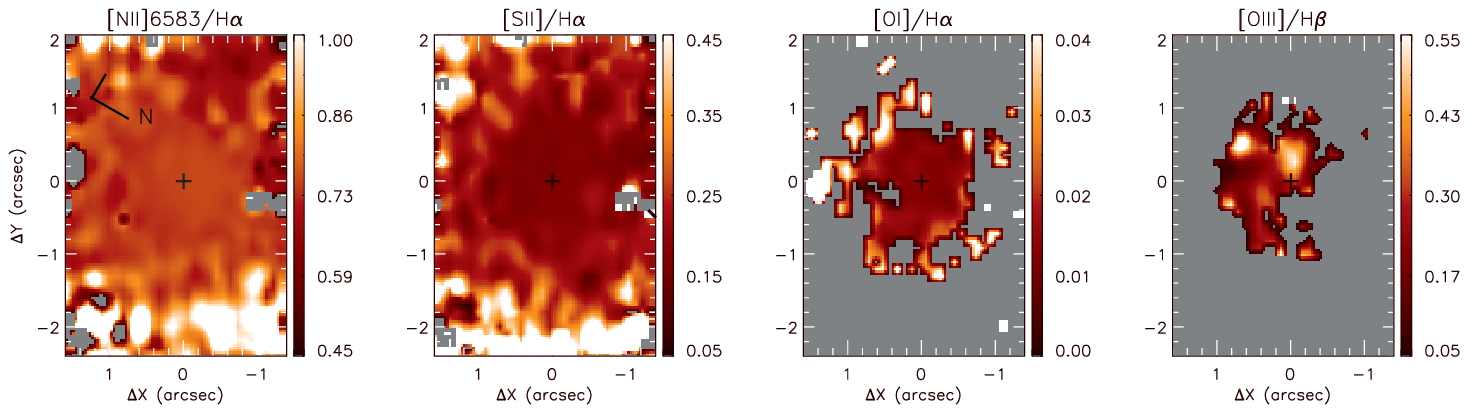}
    \caption{Emission-line ratio maps for IRAS03056. From left to right: [N\,{\sc ii}]$\lambda$6583/H$\alpha$, [S\,{\sc ii}]$\lambda$6717/H$\alpha$,  [O\,{\sc i}]$\lambda$6300/H$\alpha$  and [O\,{\sc iii}]$\lambda$5007/H$\beta$ flux-ratio maps.  Grey regions correspond to locations where the signal-to-noise ratio was not high enough to measure one or both lines of each ratio map.}
    \label{fig:ratio}
\end{figure*}

The [N\,{\sc ii}]$\lambda$6583/H$\alpha$, [S\,{\sc ii}]$\lambda$6717/H$\alpha$, [O\,{\sc i}]$\lambda$6300/H$\alpha$ and [O\,{\sc iii}]$\lambda$5007/H$\beta$ emission-line flux ratios can be used to investigate the origin of the line emission. The lines of each ratio above are close in wavelength and thus the effects of dust extinction can be neglected. Figure \ref{fig:ratio} presents the flux ratio maps for [N\,{\sc ii}]$\lambda$6583/H$\alpha$, [S\,{\sc ii}]$\lambda$6717/H$\alpha$, [O\,{\sc i}]$\lambda$6300/H$\alpha$ and [O\,{\sc iii}]$\lambda$5007/H$\beta$.

The [N\,{\sc ii}]$\lambda$6583/H$\alpha$ ratio map is remarkably uniform, presenting small values of roughly 0.75, except for the regions located to the north and northwest of the nucleus, close to the borders of the GMOS FoV. The [S\,{\sc ii}]$\lambda$6717/H$\alpha$  map shows constant values of $\sim0.2$ surrounding the nucleus at  $\sim$1$^{\prime\prime}$, approximately  coincident with the ring of  enhanced H$\alpha$ emission seen in Figure \ref{fig:fluxo}. Some smaller values are seen within the ring and values of up to  0.5 are seen at larger distances from the nucleus. The [O\,{\sc i}]$\lambda$6300/H$\alpha$ ratio map presents constant values of 0.02 within the inner 1$^{\prime\prime}$. The [O\,{\sc iii}]$\lambda$5007/H$\beta$ ratio typically has values close to 0.2, but a few knots show higher values, reaching $\sim$ 0.5.

%\textcolor{red}{Os mapas de razões devem aparecer antes dos BBPTs. Veja a nota no caption da figura.}

\begin{figure}
\centering
	\includegraphics[width=0.8\columnwidth]{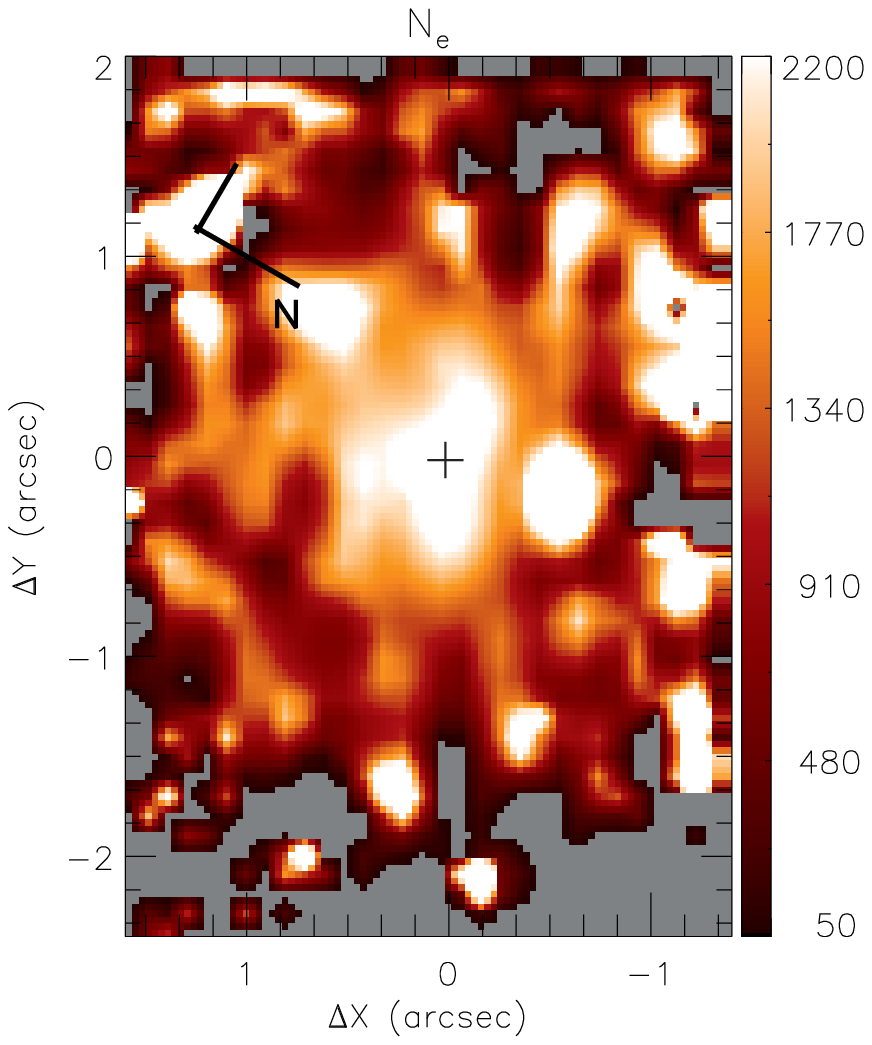}
    \caption{Electron density map for IRAS03056 obtained from the [S\,{\sc ii}] emission lines. The color bar shows the $N_e$ values in units of cm$^{-3}$.}
    \label{fig:Ne}
\end{figure}

Figure \ref{fig:Ne} shows the electron density N$_{e}$ map, measured from the [S\,{\sc ii}]$\lambda$\,6717/$\lambda$\,6731 line-ratio, assuming an electron temperature of 10\,000 K for the ionized gas as input for the {\it temden} routine in the {\sc stsdas.iraf} package. The $N_e$ map shows values ranging from 100 to 2200\,cm$^{-1}$, with the highest ones observed at the nucleus and in unresolved structures in its surroundings.

\subsection{Gas velocity fields and velocity dispersion maps}

\begin{figure*}
\includegraphics[width=1.7\columnwidth]{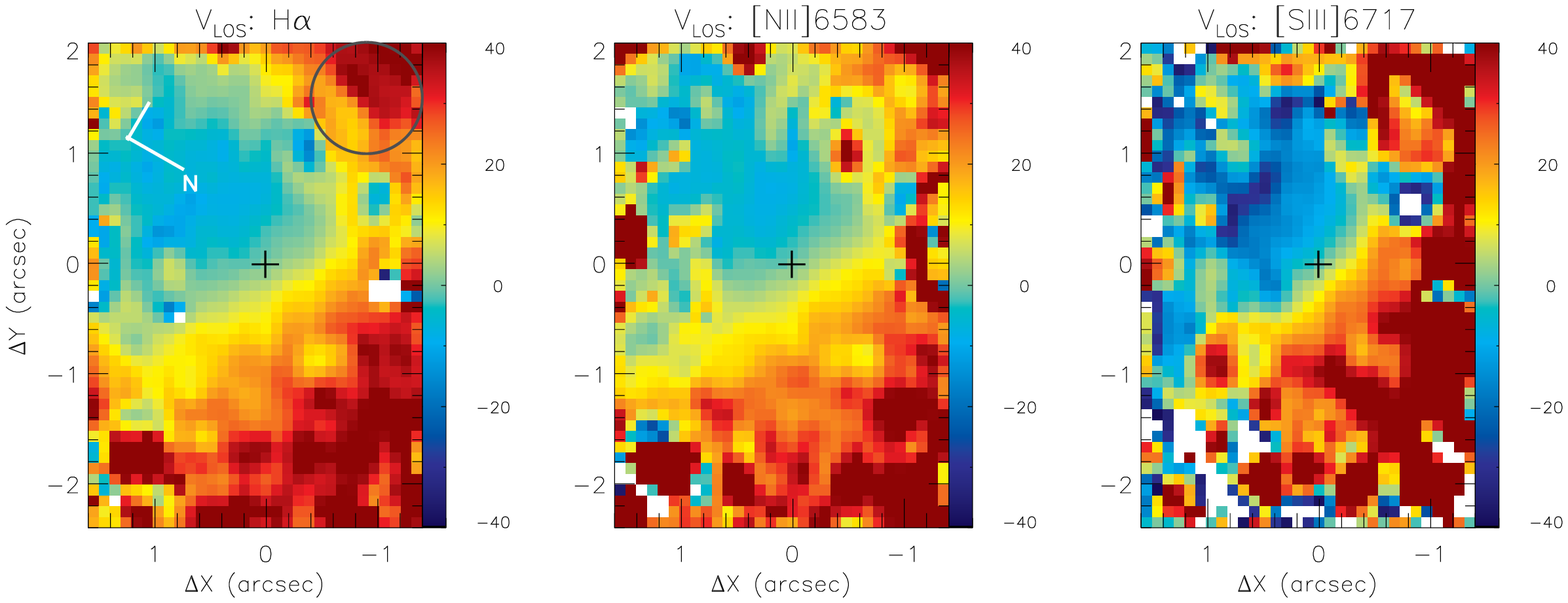}
\includegraphics[width=1.7\columnwidth]{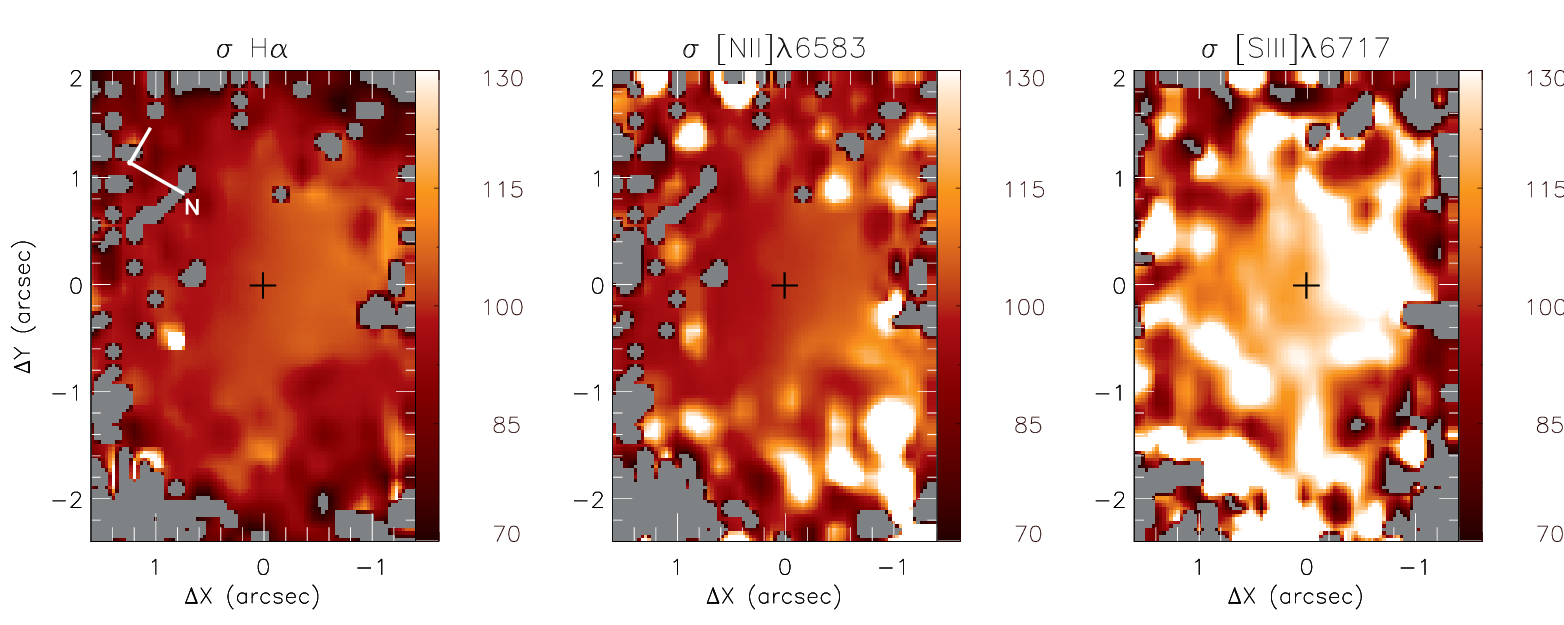}
\caption{Top panels: line-of-sight velocity fields for the H$\alpha$ (left), [N\,{\sc ii}]$\lambda$6583 (centre) and [S\,{\sc ii}]$\lambda$6717 (right) emitting gas. The color bars show the velocities in units of km\,s$^{-1}$, after the subtraction of the systemic velocity of the galaxy. Bottom panels: velocity dispersion maps for the  H$\alpha$ (left), [N\,{\sc ii}]$\lambda$6583 (centre) and [S\,{\sc ii}]$\lambda$6717 (right) emission-lines, corrected for the instrumental broadening. The color bars show the $\sigma$ values in units of km\,s$^{-1}$. The central cross in all panels marks the position of the nucleus. The circle at the top-right panel identifies the region where the inflows are detected.}
\label{fig:velocidade-sigma}
\end{figure*}

Figure \ref{fig:velocidade-sigma} presents the line-of-sight velocity fields (top panels) and the velocity dispersion maps  (bottom panels) for the H$\alpha$, [N\,{\sc ii}]$\lambda$6583 and [S\,{\sc ii}]$\lambda$6717 emission-lines. As the  H$\beta$, [O\,{\sc iii}]$\lambda$5007 and [O\,{\sc i}]$\lambda$6300 emission is seen only closer to the nucleus, and the measured kinematics is similar to that seen in the other lines showing more extended emission, we do not show the corresponding maps for these lines. The grey regions in all maps represent masked locations following the same criteria used for the flux maps, described in Sec.~\ref{flux-res}. The heliocentric systemic velocity of 8087~km s$^{-1}$ was subtracted from the observed velocities. This value was derived by modeling of the H$_{\alpha}$ velocity field with 
a disc rotation model (see section \ref{gas_kin}).

The velocity fields for all emission-lines are similar, presenting blueshifts and redshifts of $\sim40$~\kms\ to south and north, respectively. In addition, redshifts can be seen to the northeast at $\sim$2$^{\prime\prime}$ from the nucleus, close to the border of the GMOS FoV (identified by a circle in the top-right panel of Fig.~ \ref{fig:velocidade-sigma}). 

The velocity dispersion map for H$\alpha$ presents values smaller than 100~\kms\ at most locations.  The  [N\,{\sc ii}]$\lambda$6583 $\sigma$ map presents some higher values of 130~\kms\ to the north and surrounding the nucleus, while the highest $\sigma$ values are seen for the  [S\,{\sc ii}]$\lambda$6717 emitting gas, with values higher than 100~\kms\ observed at most locations. 

%\textcolor{red}{The zero velocity line presents ???? WHAT? acho que faltou concluir algo aqui} 

%\textcolor{red}{E os velocity dispersion maps???}

\section{Discussion}

\subsection{Merger stage}

%In previous studies, our group presented multiwavelength observations of the OH Megamaser galaxies IRAS23199+0123 \citep{heka18} and IRAS16399-0937 \citep{Sales2015} using similar data as presented in this work for IRAS03056. 

IRAS23199+0123 \citep{heka18} and IRAS16399-0937 \citep{Sales2015} show clear evidence of advance merger stage, as tidal tails seen in HST images. On the other hand, IRAS03056 appears to be an isolated barred spiral galaxy, as our HST images (Figs.~\ref{fig:merger} and \ref{fig:hst-images}) show no clear evidence of interaction. 

Regarding the OH maser emission, IRAS23199+0123 and IRAS16399-0937 present $L_{\rm OH}$=10$^{2.35}$ and $L_{\rm OH}$=10$^{1.7}$~L$_\odot$ respectively \citep{baan98,Darling2000}. On the other hand, IRAS03056 shows a smaller luminosity $L_{\rm OH}$=10$^{1.3}$~L$_\odot$ \citep{baan08}, suggesting that objects at more advanced merger stages present stronger OH maser emission. 

\subsection{Gas excitation and BPT diagram}

%\textcolor{red}{Primeiro falar dos 3 BPTs,para que servem? o que são as linhas? quem definiu? depois dizer como foram construidos...pixel-a-pixel,jogando medidas ruins fora (?? suspeito que possa ter algo errado nos dois últimos. O que tem no eixo y destes?) Depois falar do ponto nuclear,comparar com Baan+98 }

%\textcolor{red}{colocaria o parágrafo abaixo como uma seção, antes desta,chamada de "Star forming regions"...descreveria a estrutura circular e dai falaria das propriedades}

\citet{baldwin81} proposed three diagnostic diagrams using emission-line ratios that allow us to distinguish gas excitation by AGN or starburst activity. These BPT diagrams are based on the following four line ratios: [N\,{\sc ii}]$\lambda6583$/H$\alpha$, [O\,{\sc iii}]$\lambda5007$/H$\beta$, [S\,{\sc ii}]$\lambda$6717/H$\alpha$ and [O\,{\sc i}]$\lambda$6300/H$\alpha$. The advantage of the BPT diagrams is that emission lines of each ratio are close in wavelength and thus the extinction effects can be neglected.

\begin{figure*}
	\includegraphics[width=0.65\columnwidth]{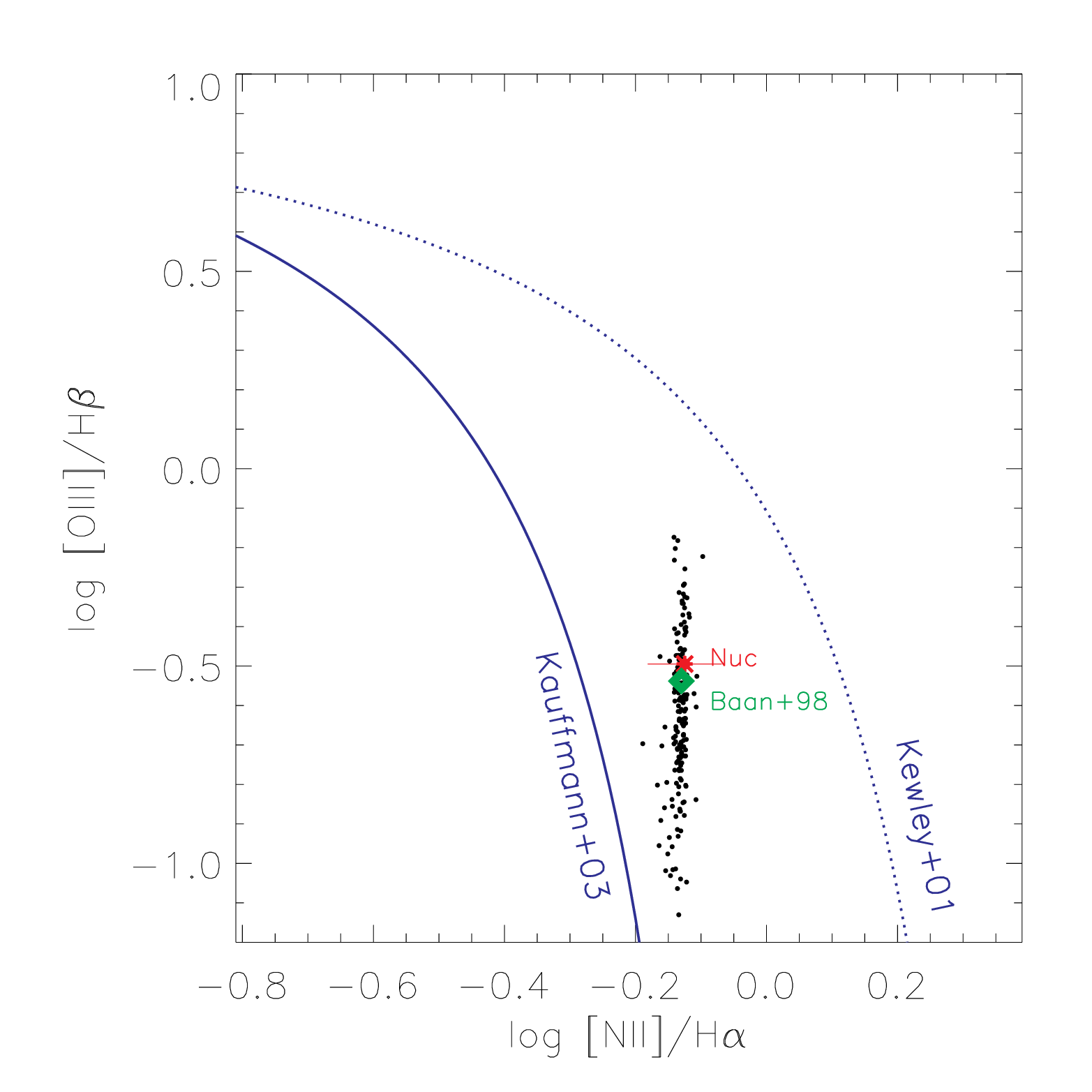}
	\includegraphics[width=0.65\columnwidth]{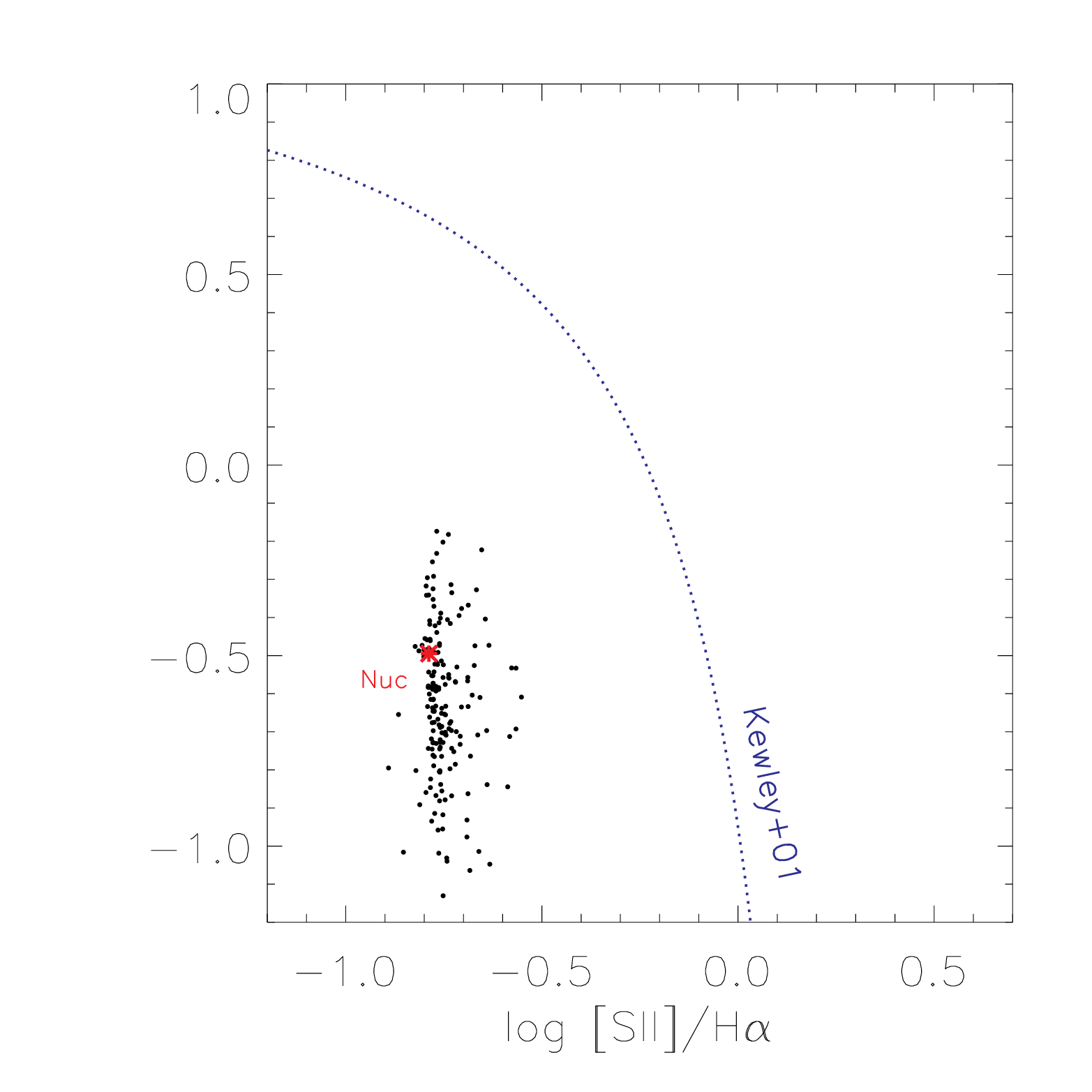}
	\includegraphics[width=0.65\columnwidth]{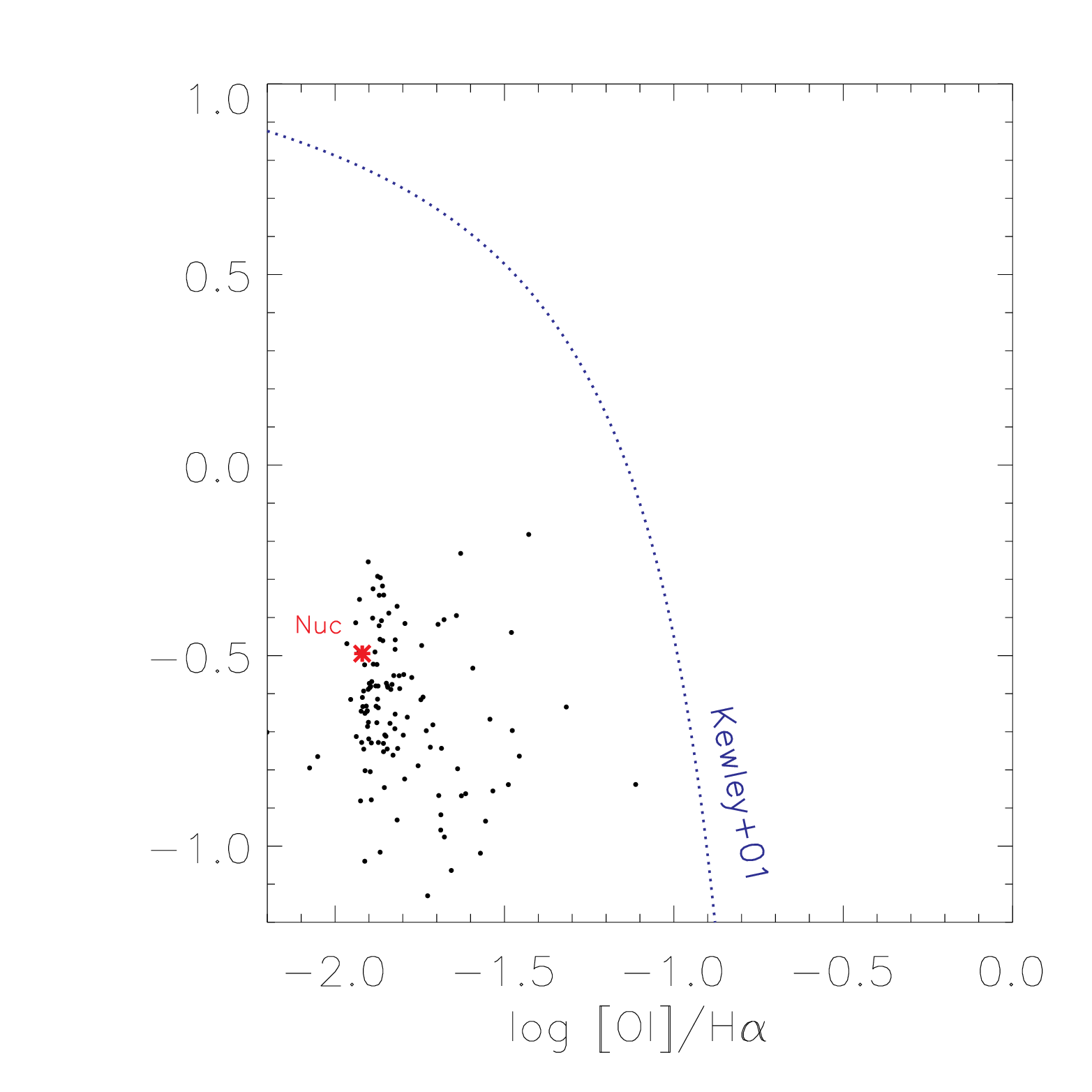}
    \caption{Left panel: [N\,{\sc ii}]$\lambda6583$/H$\alpha$ versus [O\,{\sc iii}]$\lambda5007$/H$_{\beta}$ diagnostic diagram of IRAS03056. The dotted and dashed lines represent the Kewley and Kauffmann criteria respectively. Central panel: [S\,{\sc ii}]/H$_{\alpha}$ versus [O\,{\sc iii}]$\lambda5007$/H$_{\beta}$ diagnostic diagram. Right panel: [O\,{\sc iii}]/H$_{\alpha}$ versus [O\,{\sc iii}]$\lambda5007$/H$_{\beta}$ diagnostic diagram. The red asterisks correspond to the nucleus and the green diamond corresponds to \citet{baan98} measurements}
    \label{fig:bpts}
\end{figure*}

Figure \ref{fig:bpts} presents the BPT diagrams for IRAS03056, where the x-axis show the [N\,{\sc ii}]$\lambda6583$/H$\alpha$, [S\,{\sc ii}]$\lambda$6717/H$\alpha$ and [O\,{\sc i}]$\lambda$6300/H$\alpha$ respectively and the y-axis shows the [O\,{\sc iii}]$\lambda5007$/H$\beta$ line ratios. The dotted curve corresponds to the line of \citet{kewley2001} that delineates the region occupied by AGN (to the right and above the line), while the solid line represents the revised Kauffmann criteria that separates pure star-forming galaxies (to the left an below the line) from AGN-HII composite objects \citep{Kauffmann2003}. Points located between the solid and dotted lines represent regions that have contributions to the gas excitation from both AGN and Starbursts \citep{kewley2006}.
%The BPT diagnostic diagrams for IRAS03056+2034 are shown in Figure~\ref{fig:bpts}.
To construct the BPT diagrams shown in Fig.~\ref{fig:bpts}, we calculated the flux ratio for each spaxel, excluding flux measurements with uncertainties larger than 30\,\%.

%The left panel of Fig. \ref{fig:bpts} shows the [N\,{\sc ii}]$\lambda6583$/H$\alpha$ versus [O\,{\sc iii}]$\lambda5007$/H$\beta$ diagnostic diagram, the centre panel displays the [S\,{\sc iii}]$\lambda$6717/H$\alpha$ versus [O\,{\sc iii}]$\lambda5007$/H$\beta$ diagnostic diagram and the right panel shows the [O\,{\sc i}]$\lambda$6300/H$\alpha$ versus [O\,{\sc iii}]$\lambda5007$/H$\beta$ diagnostic diagram. 

The [N\,{\sc ii}]$\lambda6583$/H$\alpha$ vs. [O\,{\sc iii}]$\lambda5007$/H$\beta$ diagnostic diagram (left panel of Fig. \ref{fig:bpts}) shows all observed points in the region between the \citet{kewley2001} and \citet{Kauffmann2003} lines, suggesting a contribution of both AGN and star-formation to the gas excitation. The other two diagrams show observed line ratios below the \citet{kewley2001} criteria for all locations.

In Fig.~\ref{fig:bpts}, the red asterisk corresponds to the nuclear position of the galaxy, as  obtained from the measurements of the emission-line ratios within a circular aperture of 0\farcs5 radius. The green diamond corresponds to the line ratios measured by \citet{baan98}, which are similar to our nuclear values. These authors performed optical classification of 42 galaxies using spectroscopic data obtained at the Palomar Observatory and classified the nucleus of IRAS03056+2034 as starburst. However, their spectra were obtained with a long-slit of width of 2$^{\prime\prime}$ and extractions of $\sim$6$^{\prime\prime}$, which corresponds to 1.1$\times$2.8\,kpc$^2$ at IRAS03056 and thus includes a large fraction of circumnuclear emission. The GMOS angular resolution is about 3  times better than that provided by the aperture used by \citet{baan98}. 

Although we were able to map the [O\,{\sc iii}]$\lambda$5007 emission only for the inner 1$^{\prime\prime}$ region, we notice that the  [O\,{\sc iii}]$\lambda$5007/H$\beta$ vary with the distance from the nucleus, where the highest values are seen. On the other hand, the [N\,{\sc ii}]$\lambda6583$/H$\alpha$ line ratio values do not change. This behaviour can be interpreted as an increasing contribution of an AGN to the production of the high excitation [O\,{\sc iii}] emission at the nucleus, which might be confirmed with higher resolution observations.

%\subsection{The nature of the nuclear emission} \label{broad}

%\subsection{Gas excitation}

%%%\begin{figure}
%%%	\includegraphics[width=\columnwidth]{ratio.ps}
%%%    \caption{[N\,{\sc ii}]/H$\alpha$ flux ratio map for IRAS23199E. Grey regions correspond to masked locations where no reliable measurements are available. The green contours in the map are from the 3~cm radio image. }
%%%    \label{fig:razaoNIIporHalpha}
%%%\end{figure}

%%\begin{figure}
  %%  \includegraphics[width=\columnwidth]{niiha-sigma.ps}
%%    \caption{[N\,{\sc ii}]$\lambda$6583/\Ha\ vs. [N\,{\sc ii}] velocity dispersion.}
 %%   \label{fig:niihasigma}
%%\end{figure}

%The dotted and dashed lines represent the Kauffman \citet{kauffmann2003} and Kewley \citet{Kewley2006} criteria. The points lying below Kauffmann line represents regions with ionization due to starburst, whereas points that lie above the Kewley line represent points of the galaxy that are ionized by an AGN. The region between the Kewley and Kauffmann lines determine the composed region, where the ionization is due to both phenomena.

\subsection{Star forming regions}

The H$\alpha$ flux map (Fig.\ref{fig:fluxo}) suggests the presence of a ring of circumnuclear star forming regions located between 0\farcs7 and 1\farcs5 from the nucleus, as several unresolved knots of higher fluxes are seen at this location. 

In order to characterize the star formation in the ring, we have extracted spectra from within a ring with inner radius of  0\farcs7 and outer radius of 1\farcs5 (identified as green circles in Fig.\ref{fig:fluxo}). 
%Once we had fluxes for these two regions, we subtracted one from each other in order to obtain the flux just for the ring.

Using the H$_\alpha$/H$_\beta$ emission-line ratio and following \citet{Izabel}, we obtained an extinction of A$_{V}\sim$0.2\,mag for the ring, which can be used to correct the observed line intensities. The H$\alpha$ luminosity of the ring (corrected by reddening) is $L_{\rm H\alpha}\approx6.2\times10^{41}$ erg s$^{-1}$, which can be used to estimate the mass of ionized gas, rate of ionizing photons and star formation rate of the ring.
%which allowed us to determine another physical properties of this region. 
%\textcolor{red}{apresentar as equaçoes usadas para calcular a massa de gás ionizado e as suposições realizadas para obter esta equçao}

In order to estimate the mass of ionized gas we used \citep[e.g.][]{peterson97}:

\begin{equation}
    \frac{M_{\rm HII}}{\rm M_{\odot}} \approx 2.3 \times 10^{5} \frac{L_{41}(H{\alpha})}{n_{3}^{2}}, 
\end{equation}
where $L_{41}$(H$\alpha$) is the H$\alpha$ luminosity in units of 10$^{41}$ erg s$^{-1}$ and n$_{3}$ is the electron density ($N_e$) in units of 10$^3$\,cm$^{-3}$. 

Using the ratio between the  fluxes of the [S\,{\sc ii}] emission lines for the ring ([S\,{\sc ii}]$\lambda$\,6717/$\lambda$\,6731$\approx$0.8)  we obtain an electron density $N_e\approx1385\,$cm$^{-3}$, assuming an electron temperature $T_e=10\,000$~K. This value is larger than those typically observed in star forming regions  \citep{diaz07,oli08}.

%We have assumed $N_e=300$ \,cm$^{-3}$, which is the mean value of electron density of circumnuclear star formation regions derived from the [S\,{\sc ii}]$\lambda$\,6717/$\lambda$\,6731 intensity ratio \citep{diaz07,oli08}.  
The estimated mass of ionized gas is $M_{\rm HII}$=7.5$\times$10$^{5}$~M$_\odot$, which is similar to the values previously found for rings of circumnuclear star forming regions \citep[e.g.][]{riffel2016,riffel09,hennig2018}. 

 %\textcolor{red}{apresentar equaçoes e sugestões para o resto}
 
 We estimated the rate of ionizing photons ($Q$[H$^{+}$]) and star formation rate (SFR) under the assumption of a continuous star formation regime.
 We derived $Q[H^{+}]$ using \citet{osterbrock}:
 
 \begin{equation}
 Q[H^{+}]=\frac{\alpha_{B}L_{H_{\alpha}}}{\alpha^{EFF}_{H_{\alpha}}h\nu_{H_{\alpha}}},
\end{equation}
where $\alpha_{B}$ is the hydrogen recombination coefficient to all energy levels above the ground level, $\alpha^{EFF}_{H_{\alpha}}$ is the effective recombination coefficient for H$\alpha$, $h$ is the Planck's constant and $\nu_{H_{\alpha}}$ is the frequency of the H$\alpha$ emission line. Using $\alpha_{B}$=2.59$\times$ 10$^{13}$ cm$^{3}$s$^{-1}$ and $\alpha^{EFF}_{H_{\alpha}}$=1.17$\times$10$^{-17}$cm$^{3}$s$^{-1}$ \citep{osterbrock} we obtain:

\begin{equation}
    \left(\frac{Q[H^{+}]}{s^{-1}}\right)=1.03\times10^{12}\left(\frac{L_{H_{\alpha}}}{s^{-1}}\right).
\end{equation}

The SFR was computed using the following relation \citep{kennicutt98}:
\begin{equation}
\frac{SFR}{\rm M_\odot\,yr^{-1}}=7.9\times10^{-42}\,\frac{L_{\rm H_\alpha}}{\rm  erg\, s^{-1}} 
\end{equation}

 We found values of ionizing photons rate of $\log$ Q[H$^{+}$]$=$53.8 and star formation rate of   SFR=4.9~M$_{\odot}$ yr$^{-1}$, which are in agreement with previous reported values for circumnuclear star forming regions in nearby galaxies \citep[e.g.][]{Wold06,Galliano08,oli08,riffel09,riffel2016,heka18}.

\subsection{Gas kinematics} \label{gas_kin}

\begin{figure*}
	\includegraphics[width=1.8\columnwidth]{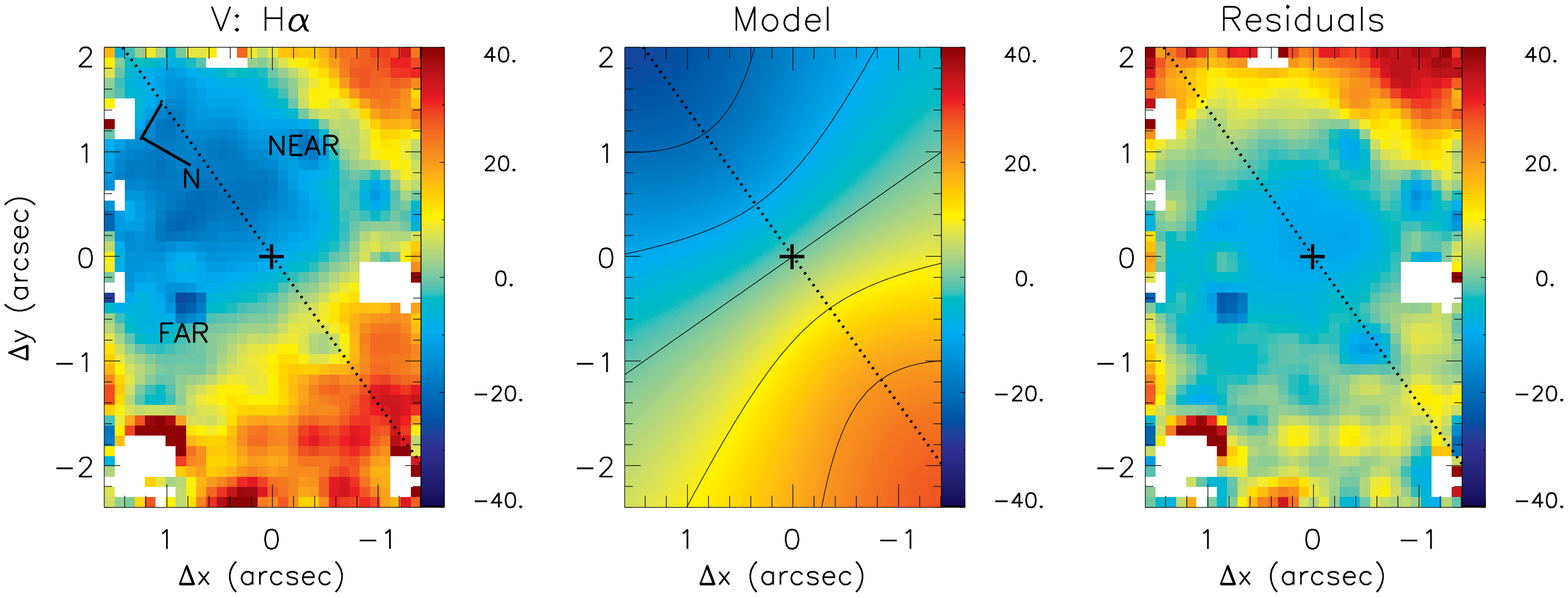}
	\caption{Observed H$\alpha$ velocity field (left), rotating disc model (middle) and residual map (right), obtained as the difference between the observed velocities and the model. 
	%The near and the far side of the galaxy disk are indicated in the left panel,
	The central cross marks the position of the nucleus, the white regions are masked locations where we were not able to fit the emission line profiles and the dotted lines represent the orientation of the line of nodes.
}
	\label{fig:rotationmodel}
\end{figure*}

The gas velocity fields of IRAS03056 presented in Fig.~ \ref{fig:velocidade-sigma} show a clear rotation pattern with blueshifts to the south and redshifts to the north of the nucleus.  In order to model these kinematics, we used a simple rotation model \citep{kruit78,bertola91}, under the assumption that the gas moves in circular orbits in the plane of the galaxy, subjected to a central gravitational potential. In this model, the rotation velocity field is given by:

\[ 
 V_{mod}(R,\psi)=v_{s}+ 
\]
\begin{equation} \label{eq-bertola}
    \frac{AR\cos(\psi-\psi_{0})\sin(i){\cos^{p}(i)}}{\{R^{2}[\sin^{2}(\psi-\psi_{0})+\cos^{2}(i)\cos^{2}(\psi-\psi_{0})]+{c_{0}}^{2}\cos^{2}(i)\}^{\frac{p}{2}}}, 
    \end{equation}
where $R$ and $\psi$ are the coordinates of each spaxel in the plane of the sky, $v_{s}$ is the systemic velocity of the galaxy, $A$ is the velocity amplitude, $\psi_{0}$ is the major axis position angle, $i$ is the disc inclination in relation to the plane of the sky ($i=0$ for face-on disc), $p$ is a model fitting parameter (for {\it p}\,=\,1 the rotation curve at large radii is asymptotically flat
while for {\it p}\,=\,3/2 the system has a finite mass) and $c_{0}$ is a concentration parameter,  defined as the radius where the rotation curve reaches 70\,\% of the velocity amplitude.

The fit of this model provides information about the physical parameters of the system, such as the systemic velocity and orientation of the kinematic major axis. Moreover, the residual map (difference between the observed velocities and the model) allow us to detect deviations from pure rotation and identify non-circular motions.

We have chosen the velocity field of the H$_{\alpha}$ emission-line to perform the fit, as this is the brightest line at most locations. We used the IDL\footnote{\rm $http://www.harrisgeospatial.com/ProductsandSolutions/Geospatial$\\$Products/IDL.aspx$} routine MPFITFUN  \citet{mark09} in order to fit the observed velocity field with Eq.~\ref{eq-bertola}.
During the fit, we excluded the redshifted region to the east of the nucleus, which clearly does not follows the rotation pattern.

%During the fit, the position of the kinematical center was fixed to the position of the peak of the continuum emission and the parameter p was fixed to $p=1.5$, as done in previous works \citep[e.g.][]{brum17,heka18}. \textcolor{red}{Foi isso mesmo? A regiao do redhift para N foi excluida do ajuste, por claramente não poder ser representada por rotação.}

The resulting best fit model is shown in the central panel of Figure  \ref{fig:rotationmodel} and its parameters are $A=40.5\pm5$ km\,s$^{-1}$, $v_{s}$=8\,087$\pm$20 km s$^{-1}$ (corrected to the heliocentric rest frame), $\Psi_{0}$=156$\pm$2$^{\circ}$, $c_{0}$=1\farcs9$\pm$0\farcs4, $i=54^\circ\pm$3$^{\circ}$. The obtained systemic velocity is about 140\,\kms\ smaller than that quoted in  NASA/IPAC Extragalactic Database (NED)\footnote{http://ned.ipac.caltech.edu}. The $\Psi_{0}$ value is consistent with the orientation of the apparent major axis of IRAS03056 as seen in the HST i-band image (Fig.~\ref{fig:hst-images}), and is about 20$^\circ$ displaced from the value quoted in NED (135$^\circ$) as obtained from 2MASS K$_{\rm S}$ photometry \citep{2mass}. The disk inclination is consistent with the value shown in NED, obtained from the 2MASS photometry.   

Figure \ref{fig:rotationmodel} also presents the observed H$_{\alpha}$ velocity field in the left panel and residuals map in the right panel. The velocity field presents redshifts to the north and blueshifts to the south, which are well described by the rotating disk model,  as indicated by the residual map, that shows values at most locations that are within -10 \kms and +10 \kms, except for a region to the east of the nucleus (close to the border of the GMOS FoV, identified as a circle in Fig.~\ref{fig:velocidade-sigma}), whose kinematics is not reproduced by our model. At the same location, the HST H$\alpha+$[N\,{\sc ii}] narrow-band image (Figure \ref{fig:hst-images}) shows a strip of enhanced emission, which may be due to a patchy spiral arm.

Considering the observed velocity fields of Fig.~\ref{fig:velocidade-sigma} and the orientation of the spiral arms seen in the bottom-left panel of Fig.~\ref{fig:hst-images}, we conclude that the northeast is the near side of the galaxy disk, while the southwest is the far side of the galaxy, under the assumption that the spiral arms seen is IRAS03056 are trailing. This assumption is supported by the fact that the vast majority of galaxies present spiral arms of the trailing type \citep[e.g.][]{binney08}. Thus, the excess of redshifts seen to the east of the nucleus can be interpreted as being due to gas streaming motions along a nuclear spiral arm, under the assumption that the gas is located at the plane of the galaxy. Similar gas inflows along nuclear spiral arms have been previously observed in nearby active galaxies \citep[e.g.][]{n4051,mrk79,sanchez09,ven10,fathi13,sm7213,luo16,busch17}.

%\subsection{Estimating the Mass Accretion Flow}

%Figure \ref{fig:rotationmodel} presents the residual map for the rotation model fit. One can see that the gas within $\sim$ 2\farcs0 from the nucleus towards the north direction is in redshift and coincident with the orientation of a nuclear bar (see Fig. \ref{fig:hst-images}). 
We can use these residual velocities observed along the nuclear spiral arm to estimate the mass inflow rate. Assuming that the gas is streaming towards the centre, we estimate the mass inflow rate as

\begin{equation}
    M_{in}= N_{e} v \pi r^{2} m_{p} f, 
\end{equation}
where $N_{e}$ is the electron density, $v$ is the inflowing velocity, $m_{p}$ is the proton mass, $\pi r^{2}$ is the area through which the gas is flowing and $f$ is the filling factor. One can estimate the filling factor using

\begin{equation}
    L_{H\alpha} \sim f N_{e}^{2} J_{H\alpha}(T) V,
\end{equation}
 where $J_{H\alpha}(V)$=3.534$\times$10$^{-25}$ erg cm$^{-3}$s$^{-1}$ \citep{osterbrock} and $L_{H_{\alpha}}$ is the H$_{\alpha}$ luminosity emitted by a region with volume $V$. 
 
 We assume that the volume of the inflowing gas region can be approximated by that of a cylinder with radius $r$ and height $h$. Thus we obtain
 
 \begin{equation}
    \dot{M}=\frac{m_{p}vL_{H\alpha}}{J_{H_{\alpha}}(T)N_{e}h}\label{eqin}
\end{equation}
 
 In order to obtain L$_{H\alpha}$ and $N_{e}$ we measured the integrated flux of H$\alpha$ and the [S\,{\sc ii}] line ratio within the redshifted region seen in the residual map. The resulting H$\alpha$ flux is $F_{H\alpha}\approx6.5\times10^{-14}\,{\rm erg s^{-1} cm^{-2}}$,  which corresponds to $L_{H\alpha}\approx1.05\times10^{41}$\,erg\,s$^{-1}$ assuming the distance of 116 Mpc for IRAS03056. The mean [S\,{\sc ii}]$\lambda$\,6717/$\lambda$\,6731 ratio for the redshifted region is $\sim$0.88. Using the {\it temden} {\sc iraf} task and adopting an electron temperature of  $T_{e}$=10\,000K, we estimate  $N_{e}=988\pm100$ cm$^{-3}$. 
 
 To estimate the mass inflow rate, we assume  $h=1$\,kpc (1\farcs8), as the distance between the nucleus of the galaxy and the redshifted region (center of the circle in Fig. \ref{fig:rotationmodel}), as measured directly from our maps. The mean inflow velocity is  $v=45$\,km s$^{-1}$,  measured from the residual map and corrected for the inclination of the disk ($i=54^\circ$, as derived from the rotation model). The resulting mass inflow rate obtained from Eq.~\ref{eqin} is $\dot{M}\approx7.7\times10^{-3}\,{\rm M_\odot\,yr^{-1}}$. The determination of the uncertainty in $\dot{M}$ is not an easy task. Considering the derived uncertainties for $i$ and $N_e$, we obtain and uncertainty of $1.2\times10^{-3}\,{\rm M_\odot\,yr^{-1}}$. 
he $\dot{M}$ value derived for IRAS0356 is consistent with those of previous estimates of mass inflow rates in ionized gas for nearby Seyfert galaxies \citep[e.g.][]{sb07,sm11,sm7213,sm1667,sm1358,n5929,sanchez09,ven10}.

As a speculation, we can estimate the dynamical time ($\Delta t_{\rm in}$) it takes for the inflowing gas to move from 1~kpc (the adopted distance) to the nucleus. Assuming that the inflow has a constant velocity ($v=45$\,km s$^{-1}$), we obtain $\Delta t_{\rm in}\approx2\times10^{7}$\,yr, which is consistent with typical AGN lifetimes \citep[e.g.][]{hopkins06}.

%The mass inflow rate can be compared to the mass accretion rate ($\dot{m}$) onto the central SMBH of IRAS03056, which can be obtained by 
%\begin{equation}
%\dot{m} = \frac{L_{\rm bol}}{c^2\eta},
%\label{txacre}
%\end{equation}
%where $\eta$ is the efficiency of conversion of matter into energy, 
%and $L_{\rm bol}$ is the AGN bolometric luminosity and $c$ is the speed of light. The bolometric luminosity can be estimated from $L_{\rm bol}=3500\,L_{\rm OIII}$, where $L_{\rm OIII}$ is the luminosity of the [O\,{\sc iii}]$\lambda5007$ emission line \citep{heckman04}. \textcolor{red}{We measure $L_{\rm OIII}=XXXXXXXXXXXX {\rm erg s^{-1}}$ within a nuclear aperture of 0\farcs5 radius, resulting in  $L_{\rm bol}= {\rm erg s^{-1}}$. Using this value and adopting $\eta=0.1$, we obtain ---puts, acho que não faz sentido esta conta, ja que não temos certeza de ter agn }

\section{Conclusions}

We have analyzed Gemini GMOS-IFU, VLA and HST data of the galaxy IRAS03056+2034, which is known to host a source of OH megamaser emission. The GMOS observations cover the inner 1.7$\times$2.5~kpc$^2$ at a spatial resolution of 506 pc and velocity resolution of 90 km\,s$^{-1}$. Our main conclusions are:

%Our observations cover the inner 9.5$\times$13.5~kpc$^2$ of the galaxy at a spatial resolution of 2.3~kpc and velocity resolution of $\sim70$\kms.  In addition, we present HST and radio images, which are compared to the emission-line flux distributions and kinematics derived from the IFU data. Our main conclusions are:

\begin{itemize}

\item The HST images reveal flocculent spiral arms, evidencing several knots of emission  along them. Comparing the GMOS-IFU flux distributions with the [N\,{\sc ii}]+H$_\alpha$ HST image we associate these knots with star forming regions located within a ring with inner radius of 337 pc and outer radius of 786 pc.

\item The 6 and 20-cm VLA image show compact radio emission at the nucleus of IRAS03056. No extended emission is observed.

\item Considering the H$_\alpha$ flux of the ring of circumnuclear star forming regions we derived that it has  mass of ionized gas of 7.5$\times$10$^{5}$ $M_{\odot}$, ionized photons rate of log $Q[H^{+}]$=53.8, and star formation rate of 4.9 M$_{\odot}$ yr$^{-1}$.

\item Based on emission-line ratios, we conclude that the nucleus of IRAS0356 shows line ratios consistent with the presence of both an AGN and starburst activity.

\item The electron density derived from the [S\,{\sc ii}]$\lambda6719/\lambda6731$ lines reaches values of up to 2200~cm$^{-3}$ in the nucleus of the galaxy.

\item The gas velocity fields show a rotation pattern with the south side of the disk approaching and the north side receding. The observed projected velocity amplitude is $\sim$40\,km\,s$^{-1}$ and the kinematic major axis is oriented along $\Psi_0=156^\circ$. The gas kinematics is well reproduced by a disk rotating model, assuming circular orbits in the galaxy plane.

\item Besides the rotating disk component, the gas kinematics reveal also an excess of redshifts seen at 1\farcs8 east of the nucleus with velocity of 45~km\,s$^{-1}$. This component which is possibly associated with a patchy nuclear spiral arm seen in the HST image, is located at the far side of the galaxy, and can be interpreted as inflows towards the nucleus. In this case, we estimate an ionized gas mass inflow rate of $7.7\times10^{-3}$M$_\odot$ yr$^{-1}$, which is similar to the inflow rate observed in nearby Seyfert galaxies.

\end{itemize}

\section*{Acknowledgements}
We thank the referee for his/her suggestions that improved the paper. 
This work is based on observations obtained at the Gemini Observatory, 
which is operated by the Association of Universities for Research in Astronomy, Inc., under a cooperative agreement with the 
NSF on behalf of the Gemini partnership: the National Science Foundation (United States), the Science and Technology 
Facilities Council (United Kingdom), the National Research Council (Canada), CONICYT (Chile), the Australian Research 
Council (Australia), Minist\'erio da Ci\^encia e Tecnologia (Brazil) and south-eastCYT (Argentina). Support for program HST-SNAP 11604 was provided by NASA through a grant from the Space Telescope Science Institute, which is operated by the Association of Universities for Research in Astronomy, Inc., under NASA contract NAS 5-26555.
This research has made use of the NASA/IPAC Extragalactic Database (NED) which is operated by the Jet Propulsion Laboratory, California Institute of Technology, under contract with the National Aeronautics and Space Administration.
We acknowledge the usage of the HyperLeda database (http://leda.univ-lyon1.fr).
C. H. thanks for CAPES financial support.  RAR and DAS acknowledge support from CNPq and FAPERGS.

%%%%%%%%%%%%%%%%%%%%%%%%%%%%%%%%%%%%%%%%%%%%%%%%%%



%\bibliographystyle{mnras}


%%%%%%%%%%%%%%%%%%%%%%%%%%%%%%%%%%%%%%%%%%%%%%%%%%

%%%%%%%%%%%%%%%%% APPENDICES %%%%%%%%%%%%%%%%%%%%%

%\appendix



%%%%%%%%%%%%%%%%%%%%%%%%%%%%%%%%%%%%%%%%%%%%%%%%%% 0.88 resolução angular.

% Don't change these lines
\bsp	% typesetting comment
\label{lastpage}
\end{document}